\documentclass[10pt,twocolumn,twoside]{IEEEtran}
\IEEEoverridecommandlockouts

\usepackage{cite}
\usepackage{amsmath,amssymb,amsfonts}
\usepackage{algorithmic}
\usepackage{graphicx}
\usepackage{textcomp}
\usepackage{balance}
\usepackage{xcolor}
\usepackage[utf8]{inputenc}
\usepackage{algorithm}
\usepackage{algorithmic}
\usepackage{enumitem}
\usepackage{subcaption}

\def\BibTeX{{\rm B\kern-.05em{\sc i\kern-.025em b}\kern-.08em
    T\kern-.1667em\lower.7ex\hbox{E}\kern-.125emX}}
\usepackage{lipsum}

\title{\LARGE \bf 
Deadline-Aware Electric Vehicles Charging with Distribution Transformer Overload Mitigation
}

\author{B Hari Kiran Reddy
\vspace{-0.5cm}
 \thanks{ }
}

\begin{document}
\begingroup
\allowdisplaybreaks

\maketitle

\begin{abstract}
High adoption of electric vehicles (EVs) can overload distribution transformers when charging requests with heterogeneous departure deadlines compete for limited capacity. Most existing coordination schemes enforce hard deadlines and strict transformer limits, implicitly assuming feasibility and failing under severe congestion. We propose a deadline-aware EV charging framework that explicitly trades off transformer thermal aging and charging service quality under capacity-constrained operation. We model transformer stress using a convex aging proxy and soften charging deadlines via penalty-weighted unmet energy at departure. We further develop a low-complexity online charging policy that prioritizes EVs based on a marginal-cost-aware urgency index. We demonstrate through case studies under increasing EV penetration that the proposed approach reduces transformer aging while preferentially allocating limited capacity to time-critical EVs, closely approximating offline benchmark performance using only real-time information.
\end{abstract}

\begin{IEEEkeywords}
Electric vehicle charging, distribution transformer aging, soft deadline scheduling, online optimization, asset-aware demand management
\end{IEEEkeywords}


\section{Introduction}

The rapid growth of EV adoption is placing increasing stress on distribution-level infrastructure, particularly residential and commercial transformers that were not originally designed for large, coincident charging loads. While EV charging is inherently time-flexible, clustering of arrivals and heterogeneous departure deadlines can lead to short-duration but severe overloads, accelerating transformer insulation aging and increasing the risk of premature asset failure \cite{singh2019distribution, arumugam2021failure}. As EV penetration continues to rise, utilities face a fundamental operational trade-off between protecting distribution assets and meeting customer charging expectations.

A substantial body of literature has investigated coordinated EV charging to mitigate distribution network constraints, including feeder congestion, voltage violations, and transformer overload \cite{gan2013optimal, sortomme2011optimal, ma2013decentralized}. Most existing approaches enforce EV charging deadlines and network limits as hard constraints, implicitly assuming that sufficient capacity exists to satisfy all demands. Under high EV penetration or constrained infrastructure, however, such formulations become infeasible and fail to describe realistic operating conditions in which not all charging requests can be fully served.

To address this issue, several transformer-aware charging strategies have been proposed that strictly cap transformer loading to avoid overload \cite{deilami2011real, chen2012autonomous}. While effective at asset protection, these approaches typically curtail charging uniformly or proportionally across EVs, without accounting for heterogeneous urgency or service priorities. As a result, vehicles with imminent departures or higher operational importance may experience significant unmet charging demand, even when limited flexibility remains.

In contrast, soft-deadline scheduling frameworks, widely studied in communication networks and real-time computing, allow controlled service degradation when system resources are insufficient, by penalizing deadline violations rather than forbidding them outright \cite{georgiadis2006resource, andrews2001scheduling}. Despite their conceptual relevance, such soft-deadline formulations have seen limited adoption in distribution-level EV charging, particularly in conjunction with asset-health metrics such as transformer thermal aging. Therefore, we study the following open critical question:
\textit{How should limited distribution transformer capacity be allocated among EVs with heterogeneous departure deadlines when it is impossible to satisfy all charging requests without overloading the transformer?}

In this paper, we propose a deadline-aware EV charging framework that explicitly captures the operational trade-off between transformer thermal aging and charging service quality under congestion. We first formulate the problem as a deterministic offline optimization and as a stochastic Markov decision process (MDP), which provides a principled characterization of transformer-constrained EV charging with uncertain arrivals and heterogeneous deadlines. Transformer stress is penalized using a convex overload proxy consistent with standard aging models, while charging deadlines are softened through penalty-weighted unmet energy at departure, preserving feasibility under high EV penetration and enabling graceful degradation. Leveraging the structural insights revealed by the MDP formulation, we then develop a low-complexity online charging policy that allocates limited capacity based on penalty-weighted urgency and marginal aging costs using only local, real-time information. Finally, we conduct a series of controlled case studies that stress transformer capacity under varying congestion and flexibility conditions, illustrating how different charging policies navigate the asset–service trade-off and demonstrating that the proposed online policy closely approximates offline benchmark behavior without requiring future information.


\section{Offline Problem Formulation}
\label{sec:off_prb_frm}

We consider a distribution transformer supplying a population of EVs over a finite time horizon
$\mathcal{T} := \{1,2,\dots,T\}$ with interval length $\Delta$.
In this section, we assume full knowledge of EV arrivals, departures,
charging demands, and non-EV base load, and formulate a deterministic
optimization problem that serves as an offline benchmark.
Although such complete information is not available in practice, the
offline formulation characterizes the fundamental trade-off between
transformer aging and charging service quality and provides a reference
against which online policies can be evaluated.

\subsection{EV Model}

Let $\mathcal{J}$ denote the set of EVs present over the horizon.
Each EV $j \in \mathcal{J}$ arrives at time $a_j$, departs at a deadline
$d_j$, requires total energy $E_j$ (kWh), and has a maximum charging
power $\bar{p}_j$ (kW). The decision variable $p_{j,t} \ge 0$ denotes the
charging power assigned to EV $j$ at time $t$.
Charging is permitted only within the availability window, i.e.,
$p_{j,t} = 0$ for all $t \notin [a_j,d_j]$.

The total energy delivered to EV $j$ before departure is
$E_j^{\mathrm{del}} := \sum_{t=a_j}^{d_j} p_{j,t} \Delta$.
Any unmet energy at departure is
$\delta_j := (E_j - E_j^{\mathrm{del}})^+$, where $(x)^+ := \max\{x,0\}$
denotes the positive-part operator.
Rather than enforcing hard charging deadlines, we adopt a soft deadline
model in which unmet energy incurs a penalty.
 
Each EV is associated with
a coefficient $\pi_j > 0$ that reflects the relative severity of leaving
EV $j$ undercharged. Importantly, $\pi_j$ does not represent monetary
cost or user importance, instead, it encodes \emph{charging flexibility}.
Larger values of $\pi_j$ correspond to EVs with limited temporal
flexibility (e.g., commuters with fixed morning departures or fleet
vehicles with scheduled usage), while smaller values represent more
flexible charging sessions (e.g., overnight residential charging or
destination charging where partial completion is acceptable).
This formulation enables explicit service differentiation and ensures
feasibility when transformer capacity is insufficient to satisfy all
demands.

\subsection{Transformer Loading and Aging Model}

Let $L_t^{\mathrm{base}}$ denote the non-EV load at time $t$ and
$P_t^{\mathrm{EV}} := \sum_{j \in \mathcal{J}} p_{j,t}$ the aggregate EV
charging load. The total transformer loading is
$P_t := L_t^{\mathrm{base}} + P_t^{\mathrm{EV}} $.

The transformer has rated capacity $P^{\mathrm{rated}}$ and an
operational limit $P^{\max}$. Thermal stress is modeled using the convex
overload proxy
$\Phi(P_t) := \left( \frac{P_t}{P^{\mathrm{rated}}} \right)^{\alpha}$, where $\alpha > 1$,
which captures the rapid increase in aging as loading approaches rated
capacity and is consistent with standard transformer aging models
\cite{ieee_c57110, swift_transformer_aging}. Rather than directly modeling hotspot temperature and insulation degradation, this proxy is interpreted as a measure of \emph{equivalent aging} relative to nominal operation. 
Specifically, the cumulative quantity $\sum_t \Phi(P_t)\Delta$ represents the amount of aging accrued over the horizon, which can be normalized to express aging in terms of equivalent operating days or years under rated loading. 
This interpretation enables a physically meaningful comparison of transformer lifetime consumption across different charging policies while retaining computational tractability.

\subsection{Optimization Problem}

The offline problem can now be formulated as:
\begin{subequations}
\begin{align}
\min \quad
& \sum_{t \in \mathcal{T}} \Phi(P_t)
+ \sum_{j \in \mathcal{J}} \pi_j \, \frac{\delta_j}{E_j}
\label{obj} \\[3pt]
\text{s.t.} \quad
& 0 \le p_{j,t} \le \bar{p}_j,
&& \forall j \in \mathcal{J},\ \forall t \in \mathcal{T}, \\
& p_{j,t} = 0,
&& \forall j \in \mathcal{J},\ \forall t \notin [a_j,d_j], \\
& L_t^{\mathrm{base}} + \sum_{j \in \mathcal{J}} p_{j,t}
\le P^{\max},
&& \forall t \in \mathcal{T}, \\
& \delta_j \ge E_j - \sum_{t=a_j}^{d_j} p_{j,t} \Delta,
&& \forall j \in \mathcal{J}, \\
& \delta_j \ge 0,
&& \forall j \in \mathcal{J}.
\end{align}
\end{subequations}

The objective in \eqref{obj} captures the trade-off between transformer
thermal stress and charging service quality. The normalization of unmet
energy by $E_j$ ensures comparability across EVs with heterogeneous
demands and prevents bias toward larger batteries. Unlike formulations
with hard deadlines, this model allows controlled service degradation
under congestion, reflecting realistic operating conditions in which
some charging shortfall may be unavoidable.

\section{MDP Formulation}
\label{sec:mdp}

We now extend the offline formulation of Section~\ref{sec:off_prb_frm} to an online stochastic setting.
Uncertainty arises from EV arrivals, departures, energy demands, and base load evolution.
The system is modeled as a finite-horizon MDP,
where charging decisions are made causally based on available information.

\subsection{State}

At time $t$, the system state is
\[s_t := \Big( L_t^{\text{base}},\ \{(e_{j,t}, d_j, \bar p_j)\}_{j \in \mathcal{J}_t} \Big),\]
where $\mathcal{J}_t$ is the set of connected EVs and $e_{j,t}$ is the remaining energy demand of EV $j$.
The remaining energy evolves deterministically as
$e_{j,t+1} = \big(e_{j,t} - p_{j,t}\Delta\big)^+$.

\subsection{Action}
The control action at time $t$ is the vector of EV charging powers, $a_t := \{p_{j,t}\}_{j \in \mathcal{J}_t}$,
chosen from the feasible action set $\mathcal{A}(s_t)$ defined by transformer and charger limits:
\begin{align*}
\mathcal{A}(s_t) := \Big\{ \{p_{j,t}\}_{j\in\mathcal{J}_t}\ :\ 
& 0 \le p_{j,t} \le \bar p_j,\ \forall j\in\mathcal{J}_t,\\
& L_t^{\text{base}} + \sum_{j\in\mathcal{J}_t} p_{j,t} \le P^{\max} \Big\}.
\end{align*}

\subsection{Exogenous Information and Transition Kernel}
Let $w_{t+1}$ denote exogenous information revealed after the action at time $t$ is chosen.
Specifically, $w_{t+1}$ includes:
(i) the evolution of the non-EV base load $L_{t+1}^{\text{base}}$, and
(ii) newly arriving EVs $\mathcal{J}_{t+1}^{\mathrm{arr}}$ with parameters $(E_j,d_j,\bar p_j)$. Accordingly, the set of connected EVs evolves as
\[
\mathcal{J}_{t+1}
=
\{ j \in \mathcal{J}_t : d_j \ge t+1 \}
\cup
\mathcal{J}_{t+1}^{\mathrm{arr}}.
\]


The system state evolves according to
\[
s_{t+1} = f(s_t,a_t,w_{t+1}),
\]
where $f(\cdot)$ represents the deterministic state update induced by the charging action,
combined with stochastic arrivals and base-load evolution.
This induces the controlled transition kernel
$s_{t+1} \sim \mathbb{P}(\cdot \mid s_t,a_t)$.

\subsection{One-Step Cost}

The one-step cost follows from the offline formulation in Section~II and comprises
transformer aging and deadline-violation penalties. Specifically,
\[
c(s_t,a_t)
= \Phi(P_t)
+
\sum_{j \in \mathcal{J}_{t}^{\mathrm{dep}}} \pi_j\, e_{j,t} / E_{j},
\]
where $P_t = L_t^{\text{base}} + \sum_{j \in \mathcal{J}_t} p_{j,t}$ and
$\mathcal{J}_{t}^{\mathrm{dep}} := \{ j \in \mathcal{J}_t : d_j = t \}$ denotes the set of EVs departing at time $t$.

\subsection{Objective}
We seek a causal policy $\pi := \{\mu_t\}_{t=1}^T$ where $a_t = \mu_t(s_t)$ that minimizes the expected cumulative cost:
\[
\min_{\pi} \ \mathbb{E}^{\pi}\!\left[\sum_{t=1}^{T} c(s_t,a_t)\right],
\qquad \text{s.t. } a_t \in \mathcal{A}(s_t)\ \ \forall t.
\]

\subsection{Dynamic Programming Recursion}
The optimal value function satisfies the finite-horizon Bellman recursion:
\begin{align*}
&V_{T+1}(s) := 0,\\
&V_t(s_{t}) := \min_{a_{t} \in \mathcal{A}(s_{t})} \Big\{ c(s_{t},a_t) + \mathbb{E}\!\left[V_{t+1}(s_{t+1}) \mid s_{t},a_{t}\right] \Big\}, \forall t.
\end{align*}
An optimal policy is obtained by selecting any minimizer of the right-hand side at each state and time. This MDP formulation captures the stochastic and online nature of transformer-constrained EV charging. However, solving the Bellman recursion exactly is computationally intractable due to the high-dimensional and variable-cardinality state induced by random EV arrivals and heterogeneous deadlines. We therefore propose a low-complexity online policy that serves as an approximate solution to the MDP.

\section{Marginal-Cost-Aware Online Charging Algorithm}

The objective of the proposed algorithm is to allocate limited transformer capacity among connected EVs in real time when it is not possible to satisfy all charging demands simultaneously. At each time step, the algorithm evaluates the urgency of each active EV based on its remaining energy demand, time to departure, and flexibility, and compares this urgency against the marginal cost of increasing transformer loading. Charging power is then assigned sequentially, prioritizing EVs with higher urgency while respecting charger limits and transformer capacity. When the marginal cost of additional loading exceeds the expected benefit of serving further EV demand, the algorithm curtails charging to prevent excessive transformer stress. As a result, the algorithm achieves a balanced trade-off between protecting the transformer and reducing unmet charging demand at departure using only local, real-time information. The pseudocode of the proposed algorithm is presented in Algorithm~\ref{alg:mc_urgency}. The algorithm consists of three main steps, which are described below.
\begin{algorithm}[t]
\caption{Marginal-Cost-Aware Online Charging Policy}
\label{alg:mc_urgency}
\begin{algorithmic}[1]
\REQUIRE State $s_t = \big(L^{\mathrm{base}}_t,\{e_{j,t}, d_j, \bar p_j, \pi_j\}_{j\in\mathcal{J}_t}\big)$
\STATE $P^{\mathrm{avail}}_t \leftarrow P_{\max} - L^{\mathrm{base}}_t$
\IF{$P^{\mathrm{avail}}_t \le 0$}
    \STATE \RETURN $p_{j,t}=0,\ \forall j \in \mathcal{J}_t$
\ENDIF
\STATE $P_t \leftarrow L^{\mathrm{base}}_t$
\STATE $g_t \leftarrow \Phi'\!\left(\frac{P_t}{P^{\mathrm{rated}}}\right)\frac{1}{P^{\mathrm{rated}}}$
\FOR{each $j \in \mathcal{J}_t$}
    \STATE $U_j(t) \leftarrow 
    \dfrac{(\pi_j/E_j)/(d_j - t + \epsilon)}{g_t}$
\ENDFOR
\STATE Sort $\mathcal{J}_t$ in descending order of $U_j(t)$
\FOR{each $j$ in sorted order}
    \IF{$U_j(t) < 1$}
        \STATE \textbf{break}
    \ENDIF
    \STATE $p_{j,t} \leftarrow 
    \min\!\left\{
    \bar p_j,\;
    \dfrac{e_{j,t}}{\Delta},\;
    P^{\mathrm{avail}}_t
    \right\}$
    \STATE $P^{\mathrm{avail}}_t \leftarrow P^{\mathrm{avail}}_t - p_{j,t}$
    \STATE $P_t \leftarrow P_t + p_{j,t}$
    \STATE $g_t \leftarrow \Phi'\!\left(\frac{P_t}{P^{\mathrm{rated}}}\right)\frac{1}{P^{\mathrm{rated}}}$
\ENDFOR
\STATE \RETURN $a_t = \{p_{j,t}\}_{j\in\mathcal{J}_t}$
\end{algorithmic}
\end{algorithm}

\subsection{Marginal Aging Cost of Distribution Transformer}

 The marginal aging cost associated with an incremental increase in charging power is given by
\[
g_t \;\triangleq\;
\frac{\partial}{\partial P_t}
\Phi\!\left(\frac{P_t}{P^{\mathrm{rated}}}\right)
=
\Phi'\!\left(\frac{P_t}{P^{\mathrm{rated}}}\right)
\frac{1}{P^{\mathrm{rated}}}.
\]
The quantity $g_t$ represents the incremental aging penalty per unit of additional transformer loading at time $t$ and increases rapidly as $P_t$ approaches rated capacity.

\subsection{Marginal Benefit--Cost Urgency Index}

For each connected EV $j \in \mathcal{J}_t$, let $e_{j,t}$ denote its remaining energy requirement and $d_j - t$ its remaining time until departure. Under the normalized objective, allocating additional charging power to EV $j$ yields a marginal reduction in expected service shortfall proportional to
$\frac{\pi_j}{E_j} \cdot \frac{1}{d_j - t + \epsilon}$,
where $\epsilon > 0$ avoids numerical singularities near departure.  

We define the marginal-cost-aware urgency index
\[
U_j(t)
\;\triangleq\;
\frac{\displaystyle \frac{\pi_j}{E_j}\,\frac{1}{d_j - t + \epsilon}}
{\displaystyle g_t},
\]
which can be interpreted as a benefit--cost ratio comparing the marginal service benefit of charging EV $j$ against the marginal transformer aging cost incurred by additional loading.

\subsection{Online Charging Policy}

At each time step $t$, the policy operates as follows. First, the available transformer headroom is computed as
\[
P^{\mathrm{avail}}_t = P_{\max} - L^{\mathrm{base}}_t.
\]
If $P^{\mathrm{avail}}_t \le 0$, no EV charging is scheduled. Otherwise, the urgency index $U_j(t)$ is evaluated for all connected EVs and the vehicles are sorted in descending order of urgency. Charging power is then allocated sequentially according to this priority order. For each EV $j$, the assigned charging power is
\[
p_{j,t} = \min\!\left\{
\bar p_j,\;
\frac{e_{j,t}}{\Delta},\;
P^{\mathrm{avail}}_t
\right\},
\]
after which the remaining headroom and marginal aging cost are updated. The allocation process terminates when either the available capacity is exhausted or when the highest remaining urgency index satisfies $U_j(t) < 1$, indicating that the marginal aging cost outweighs the marginal service benefit.

\textit{Stopping criterion:} At each time step, the algorithm allocates charging power sequentially to EVs in decreasing order of urgency. The allocation process terminates when either (i) the transformer capacity constraint becomes binding, or (ii) the marginal benefit of allocating additional charging power to any remaining EV falls below the marginal cost associated with increased transformer loading. In either case, no further charging power is assigned during the current time step.

\section{Case Study}

\subsection{Simulation Setup}
We consider a single distribution transformer supplying a population of EVs over a 24-hour horizon, discretized into 15-minute intervals. The transformer has a rated capacity of $P_{\text{rated}} = 100$~kW and an operational limit of $P_{\max} = 120$~kW, allowing short-duration overloads. Transformer thermal stress is modeled using a convex overload proxy with exponent $\alpha = 2$. The non-EV base load follows a stylized residential evening-peaking profile, with peak loading occurring between 6--9~PM and occupying a significant fraction of the transformer capacity during peak hours. To model different levels of transformer congestion, the non-EV base load $L_t^{\text{base}}$ is scaled by a factor $c \in (0,1]$, i.e., $L_t^{\text{base}} \leftarrow c \, L_t^{\text{base}},$
where larger values of $c$ correspond to higher base-load intensity and reduced available transformer headroom. Fig.~\ref{fig:base_loads} shows the non-EV base load profiles under different scaling factors $c$ representing increasing levels of transformer congestion.
\begin{figure}[H]
    \centering
    \includegraphics[width=\linewidth]{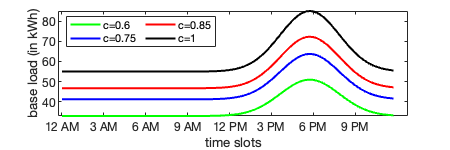}
    \caption{Non-EV base load profiles under different scaling factors $c$.}
    \label{fig:base_loads}
\end{figure}

EV arrivals are concentrated in the late afternoon and evening, consistent with residential charging behavior. Each EV requires an energy demand $E_j \in [8,30]$~kWh and has a maximum charging power $\bar p_j = 7$~kW. Parking durations and departure deadlines are heterogeneous and, under high EV penetration, insufficient to satisfy all charging demands without congestion. A subset of EVs is assigned higher penalty coefficients to represent critical users with limited charging flexibility. All results are averaged over 30 Monte Carlo days with identical base-load realizations across policies to ensure a fair comparison.


\subsection{Representative Scenarios}

To systematically evaluate charging behavior and transformer aging across increasing levels of congestion and heterogeneity, we consider four representative scenarios with progressively higher EV penetration and service differentiation.

\begin{enumerate}[label=\textbf{S\arabic*:}, leftmargin=*]

\item \textbf{Low EV Penetration (Nominal Operation).}
We simulate a lightly loaded operating condition with $|\mathcal{J}| = 8$ EVs. EV arrivals are drawn from the late-afternoon to evening period (approximately 3:00--8:00~PM), and departure deadlines are generated with a relatively long slack of $d_j - a_j \in [4,8]$~hours, providing substantial charging flexibility. The non-EV base load is scaled using $c = 0.6$, resulting in ample transformer headroom throughout the day. No EVs are designated as critical in this scenario.

\item \textbf{Moderate EV Penetration (Capacity-Constrained Operation).}
EV penetration is increased to $|\mathcal{J}| = 20$, with arrivals more tightly clustered during the evening period (approximately 2:30--7:30~PM). Charging flexibility is reduced by tightening the deadline slack to $d_j - a_j \in [2.5,6]$~hours, causing peak-hour charging decisions to affect both transformer thermal stress and service quality. The non-EV base load is increased to $c = 0.75$, limiting available transformer headroom during the evening peak. No EVs are designated as critical.

\item \textbf{High EV Penetration (Congested Operation).}
This scenario represents severe congestion with $|\mathcal{J}| = 35$ EVs arriving within a narrow evening window (approximately 2:00--7:00~PM). Departure deadlines are tight, with slack $d_j - a_j \in [2,5]$~hours, making it infeasible to fully satisfy all charging demands within the available transformer capacity. The base load is further increased to $c = 0.85$, leaving insufficient headroom during peak hours and motivating the use of soft deadlines.

\item \textbf{High EV Penetration with Critical EVs.}
This scenario uses the same congestion parameters as Scenario~S3 ($|\mathcal{J}| = 35$, arrivals between approximately 2:00--7:00~PM, deadline slack of 2--5~hours, and $c = 0.85$) to enable a controlled comparison, but introduces service heterogeneity by designating a subset of EVs as critical. Critical EVs are assigned higher penalty coefficients to evaluate whether charging policies can provide service differentiation under severe congestion while balancing transformer thermal stress.

\end{enumerate}

\subsection{Results}

\subsubsection{Comparison with and without transformer aging} Fig.~\ref{tx_age_unmet} presents an offline comparison of charging outcomes with and without explicitly accounting for transformer aging. When the aging term is included in the objective, the resulting charging schedules consistently yield lower aging factors across all scenarios, reflecting reduced thermal stress on the transformer. However, as the scenario index increases, corresponding to higher base-load intensity and EV congestion, both the aging factor and the unmet charging energy increase under both policies. This highlights an inherent trade-off: while aging-aware scheduling mitigates transformer stress, increasing congestion limits charging flexibility and leads to higher levels of unmet energy.

\begin{figure}[htbp]
    \centering
    \begin{subfigure}{0.5\textwidth}
        \centering
        \includegraphics[width=\linewidth]{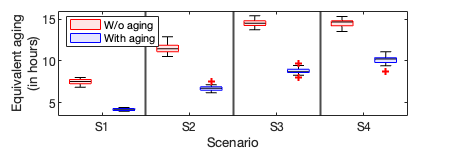}
        \caption{Equivalent transformer aging (in hours).}
    \end{subfigure}
    \begin{subfigure}{0.5\textwidth}
        \centering
        \includegraphics[width=\linewidth]{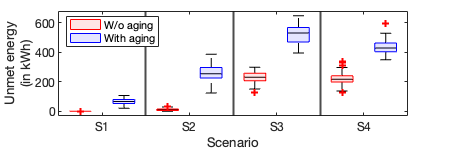}
        \caption{Total unmet EV charging energy (in kWh).}
    \end{subfigure}
    \caption{Variation of transformer aging and unmet charging energy across scenarios S1–S4, with and without explicitly accounting for transformer aging.}
    \label{tx_age_unmet}
\end{figure}

\subsubsection{Comparison of offline and online policies}
As shown in Fig.~\ref{fig:offvson}, the online charging policy results in higher equivalent transformer aging but lower unmet charging energy compared to the offline benchmark. This indicates that the online policy allocates charging power more aggressively in response to realized EV arrivals, thereby improving service completion at the expense of increased transformer loading. In contrast, the offline policy achieves lower transformer aging by leveraging full-horizon information to better manage charging schedules and avoid sustained high loading levels, albeit at the cost of higher unmet demand.

\begin{figure}[htbp]
    \centering
    \begin{subfigure}{0.5\textwidth}
        \centering
        \includegraphics[width=\linewidth]{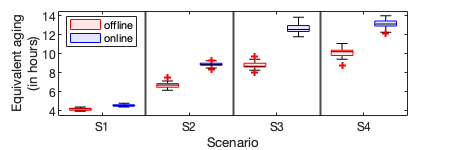}
        \caption{Equivalent transformer aging (in hours).}
    \end{subfigure}
    \begin{subfigure}{0.5\textwidth}
        \centering
        \includegraphics[width=\linewidth]{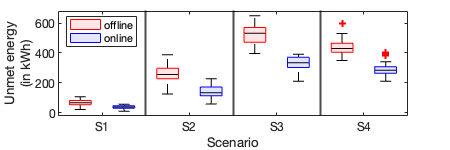}
        \caption{Total unmet EV charging energy (in kWh).}
    \end{subfigure}
    \caption{Variation of transformer aging and unmet charging energy across scenarios S1–S4 under offline and online policies.}
    \label{fig:offvson}
\end{figure}

\subsubsection{Effect of aging exponent $\alpha$}

Fig. \ref{fig:age_alpha} shows the sensitivity of transformer aging and unmet charging energy to the aging exponent $\alpha$ in the offline formulation. As  $\alpha$ increases, the aging cost becomes more convex, strongly penalizing high transformer loading. Consequently, the optimizer shifts away from peak-hour charging, resulting in a monotonic decrease in equivalent transformer aging across all scenarios. This reduction in aging is accompanied by an increase in unmet charging energy, as the scheduler increasingly curtails charging during congested periods to avoid excessive thermal stress. The trade-off is more pronounced in high-penetration scenarios (S3–S4), where limited flexibility amplifies the impact of stronger aging penalties. These results confirm that $\alpha$ directly controls the aggressiveness of asset protection in the offline benchmark, governing the balance between transformer lifetime preservation and charging service quality.

\begin{figure}[htbp]
    \centering
    \begin{subfigure}{0.5\textwidth}
        \centering
        \includegraphics[width=\linewidth]{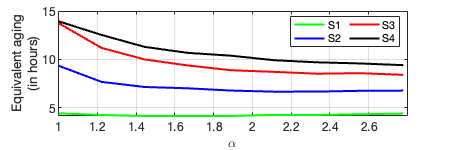}
        \caption{Equivalent transformer aging (in hours).}
    \end{subfigure}
    \begin{subfigure}{0.5\textwidth}
        \centering
        \includegraphics[width=\linewidth]{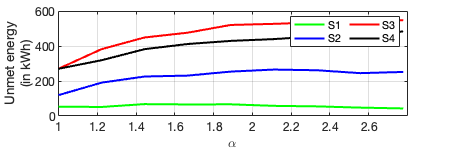}
        \caption{Total unmet EV charging energy (in kWh).}
    \end{subfigure}
    \caption{Variation of transformer aging and unmet charging energy across scenarios S1–S4 under aging exponent values.}
    \label{fig:age_alpha}
\end{figure}

\section{Conclusion}
We presented a deadline-aware EV charging framework that explicitly trades off transformer thermal aging and charging service quality under capacity-constrained operation. By softening charging deadlines and incorporating a convex aging proxy, the proposed formulation remains feasible under high EV penetration and captures realistic congestion scenarios that are not addressed by hard-deadline or hard-limit approaches. Building on this formulation, we developed a low-complexity online charging policy that allocates limited transformer capacity based on a marginal-cost-aware urgency index using only real-time information. Case study results show that the proposed policy significantly reduces transformer aging while preferentially serving time-critical EVs, and closely approximates offline benchmark performance without requiring future knowledge. These findings highlight the importance of explicitly accounting for asset health and deadline flexibility in EV charging coordination under congestion. Future work will extend the framework to feeder-level constraints, emergency overload limits, and coordination across multiple transformers.


\bibliographystyle{ieeetr}
\bibliography{references}

@standard{ieee_c57110,
  title        = {{IEEE} Guide for Loading Mineral-Oil-Immersed Transformers},
  organization = {IEEE Power \& Energy Society},
  number       = {IEEE Std C57.91-2011},
  year         = {2012}
}

@article{swift_transformer_aging,
  author  = {G. W. Swift and T. S. Molinski and W. Lehn},
  title   = {A Fundamental Approach to Transformer Thermal Modeling---Part I: Theory and Equivalent Circuit},
  journal = {IEEE Transactions on Power Delivery},
  volume  = {16},
  number  = {2},
  pages   = {171--175},
  year    = {2001}
}

@article{gan2013optimal,
  author  = {L. Gan and U. Topcu and S. H. Low},
  title   = {Optimal Decentralized Protocol for Electric Vehicle Charging},
  journal = {IEEE Transactions on Power Systems},
  volume  = {28},
  number  = {2},
  pages   = {940--951},
  year    = {2013}
}

@article{sortomme2011optimal,
  author  = {E. Sortomme and M. A. El-Sharkawi},
  title   = {Optimal Charging Strategies for Unidirectional Vehicle-to-Grid},
  journal = {IEEE Transactions on Smart Grid},
  volume  = {2},
  number  = {1},
  pages   = {131--138},
  year    = {2011}
}

@article{ma2013decentralized,
  author  = {Z. Ma and D. S. Callaway and I. A. Hiskens},
  title   = {Decentralized Charging Control of Large Populations of Plug-in Electric Vehicles},
  journal = {IEEE Transactions on Control Systems Technology},
  volume  = {21},
  number  = {1},
  pages   = {67--78},
  year    = {2013}
}

@article{deilami2011real,
  author  = {S. Deilami and A. S. Masoum and P. S. Moses and M. A. S. Masoum},
  title   = {Real-Time Coordination of Plug-In Electric Vehicle Charging in Smart Grids to Minimize Power Losses and Improve Voltage Profile},
  journal = {IEEE Transactions on Smart Grid},
  volume  = {2},
  number  = {3},
  pages   = {456--467},
  year    = {2011}
}

@article{chen2012autonomous,
  author  = {Z. Chen and L. Wu and Y. Fu},
  title   = {Real-Time Price-Based Demand Response Management for Residential Appliances via Stochastic Optimization and Robust Optimization},
  journal = {IEEE Transactions on Smart Grid},
  volume  = {3},
  number  = {4},
  pages   = {1822--1831},
  year    = {2012}
}

@article{andrews2001scheduling,
  author  = {M. Andrews and K. Kumaran and K. Ramanan and A. Stolyar and R. Vijayakumar and P. Whiting},
  title   = {Scheduling in a Queueing System with Soft Deadlines},
  journal = {IEEE Transactions on Information Theory},
  volume  = {47},
  number  = {7},
  pages   = {2740--2758},
  year    = {2001}
}

@article{georgiadis2006resource,
  author  = {L. Georgiadis and M. J. Neely and L. Tassiulas},
  title   = {Resource Allocation and Cross-Layer Control in Wireless Networks},
  journal = {Foundations and Trends in Networking},
  volume  = {1},
  number  = {1},
  pages   = {1--144},
  year    = {2006}
}

@article{singh2019distribution,
  title={{Distribution transformer failure modes, effects and criticality analysis (FMECA)}},
  author={Singh, Jaspreet and Singh, Sanjeev and Singh, Amanpreet},
  journal={Engineering Failure Analysis},
  volume={99},
  pages={180--191},
  year={2019},
  publisher={Elsevier}
}

@article{arumugam2021failure,
  title={Failure diagnosis and root-cause analysis of in-service and defective distribution transformers},
  author={Arumugam, Saravanakumar},
  journal={Energies},
  volume={14},
  number={16},
  pages={4997},
  year={2021},
  publisher={MDPI}
}
\balance

\endgroup
\end{document}